\documentclass{ws-procs9x6}

\begin{document}

\title{Exploring Particle Acceleration in Gamma-Ray Binaries}

\author{V.~Bosch-Ramon}

\address{Dublin Institute for Advanced Studies,\\
Fitzwilliam 31, Dublin 2, Ireland\\
$^*$E-mail: valenti@cp.dias.ie}

\author{F.~M.~Rieger}
\address{Max-Planck-Institut f¨ur Kernphysik,\\ 
P.O. Box 103980, 69029 Heidelberg, Germany}

\begin{abstract}
Binary systems can be powerful sources of non-thermal 
emission from radio to gamma rays. When the latter are detected, then these objects are known
as gamma-ray binaries. In this work, we explore, in the context of gamma-ray 
binaries, different acceleration processes to estimate their efficiency:  Fermi~I, Fermi~II, shear acceleration, 
the converter mechanism, and magnetic reconnection. We find that Fermi~I acceleration in a mildly 
relativistic shock can provide, although marginally, the multi-10~TeV particles required to explain observations. 
Shear acceleration may be a complementary mechanism, giving particles the final boost to reach such a high energies. 
Fermi~II acceleration may be too slow to account for the observed very high energy photons, but may be 
suitable to explain extended low-energy emission. The converter mechanism seems to require rather high 
Lorentz factors but cannot be discarded a priori. Standard relativistic shock acceleration requires a highly 
turbulent, weakly magnetized downstream medium; magnetic reconnection, by itself possibly insufficient 
to reach very high energies, could perhaps facilitate such a conditions. Further theoretical developments, and 
a better source characterization, are needed to pinpoint the dominant acceleration 
mechanism, which need not be one and the same in all sources.
\end{abstract}

\keywords{binary systems; non-thermal; gamma rays}

\bodymatter

\section{Introduction}\label{intro}

During the last decade, binary systems have turned out to be a new class of gamma-ray sources whose 
numbers are growing with the increasing sensitivity of the new instrumentation (see Refs.
\refcite{aha05a,aha05b,alb06,alb07,abd09a,abd09b,tav09a,tav09b,abd09c,sab10,abd10,bon11,cor11}). 
Different types of objects pertain to this class, like microquasars, binaries hosting a non-accreting pulsar, 
massive star binaries, and even symbiotic stars. All these sources share the characteristic of hosting powerful 
outflows that interact with themselves and their environment, and dissipate their energy partially in the form 
of non-thermal particles. Given the typical compactness of the emitter, the dynamical and radiation timescales 
are short, yielding rapidly variable emission that can reach high luminosities, and also very high energies. 
Particularly interesting in this regard are those gamma-ray binaries that reach energies $\gg$ TeV. In some 
cases, like LS~5039\cite{aha06,kha08}, the emitting particles could be as energetic as $\sim 100$~TeV, 
which poses serious constraints on any particle acceleration model that aims at explaining the observed 
radiation, as noted by Khangulyan et~al. (\refcite{kha08}). Rieger~et~al. (\refcite{rie07}) discussed different 
diffusive acceleration processes for microquasars, and also concluded that any mechanism responsible of 
the very high-energy emission should run at its limits.  In this work, we carry out a semi-quantitative analysis 
of the requirements and efficiency of diffusive acceleration processes (Fermi~I, Fermi~II and shear acceleration\cite{dru83,fer49,rie04}), the converter mechanism and magnetic reconnection (e.g. Refs. \refcite{der03,zen01}),  
for gamma-ray binaries in general. Our goal is to take a first look at the problem of extreme acceleration in 
compact galactic sources, in which dense radiation fields, together with highly supersonic sometimes relativistic 
bulk motion, shear layers, turbulence, and magnetic fields, are expected. 

\begin{figure}
\begin{center}
\psfig{file=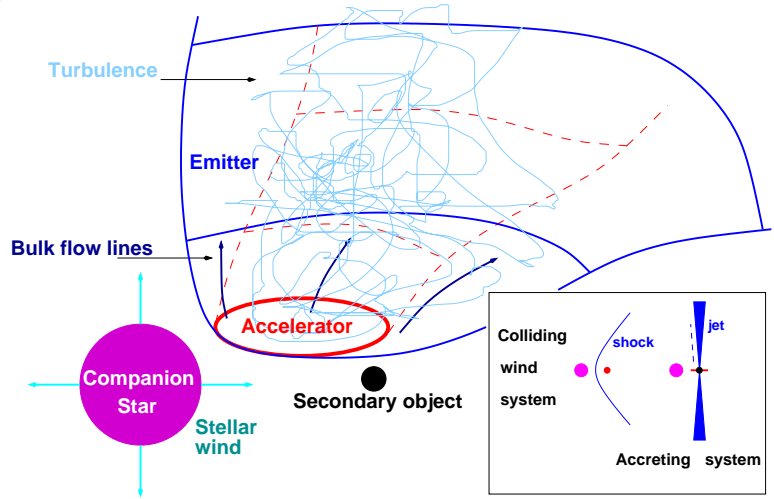,width=4in}
\end{center}
\caption{Sketch of the generic binary scenario. 
}
\label{fig1}
\end{figure}

In Figure~\ref{fig1}, a sketch of the general binary scenario discussed here is presented, showing the relevant 
elements that could play a role in the production of very energetic particles. An interaction structure is formed 
due to the presence of, for instance, a powerful relativistic outflow from a compact object (e.g. a jet or a pulsar 
wind) and a strong stellar wind. Two stellar winds could also form a similar though non-relativistic structure. 
For the case of a jet, it will be more collimated, but jet disruption may also lead to a broadening of the interaction 
region. Powerful shocks are expected to form at the colliding region: a termination shock in the pulsar, and an 
asymmetric re-collimation shock when a jet is present. In the jet scenario, internal shocks can also occur. The 
contact discontinuity between the different flows involved is subject to Rayleigh-Taylor and/or Kelvin-Helmholtz 
instabilities (neglecting the role of the magnetic field), which will trigger turbulence downstream the flow, as well 
as mixing from the two different media. Despite of its complexity, the picture can be approximately analyzed (see, 
e.g., Refs. \refcite{per08,bos11,oka11}), and different acceleration site and mechanism candidates can be 
proposed. In the presence of strong shocks, diffusive acceleration could be the dominant mechanism. For 
magnetized, highly turbulent and diluted media Fermi~II could be at work, and if strong velocity gradients are 
present, shear may occur as well. For highly relativistic and radiation or matter dense environments the converter 
mechanism could play a role, whereas magnetic reconnection could take place under the presence of strong
irregular magnetic fields. The modeling of observations tends to favor leptonic models (e.g. Refs. 
\refcite{bos06,kha07,kha08,dub08,sie08,zab11}), although hadronic models cannot be discarded (e.g. Ref. \refcite{rom03}; 
see also 
Ref. \refcite{bos09} and references therein). We will focus here mainly on electron acceleration, strongly affected at the highest 
energies by synchrotron losses, but some of our conclusions apply to protons as well. 

\section{Diffusive acceleration mechanisms}

Diffusive shock or Fermi~I acceleration takes place through repeated particle bouncing upstream-downstream 
of a shock front. The particle deflection is mediated by magnetic field irregularities of the background plasma. In 
each cycle, the particle energy gain is $\Delta E/E\propto (v_{\rm s}/c)$, where $v_{\rm s}$ is the shock velocity. 
For a mildly relativistic shock and in the Bohm diffusion limit, the acceleration timescale is $t_{\rm acc}\sim 2\pi\,
(c/v_{\rm s})^2 \,E/qBc\approx 3\,(0.5\,c/v_{\rm s})^2\,E_{\rm TeV}\,B_{\rm G}^{-1}$~s, where $E_{\rm TeV}$ is the 
particle energy in TeV and $B_{\rm G}$ the magnetic field in Gauss; $v_{\rm s}\sim 0.5\,c$ would be the
validity limit of the non-relativistic assumption. 
For reasonable radiation and matter densities, in a finite size {\it homogeneous} 
accelerator, dominant particle/energy losses are diffusion escape, and adiabatic and synchrotron cooling, with 
typical timescales $t_{\rm diff}=R^2/2\,D=15000\,R_{12}^2\,B_{\rm G}\,E_{\rm TeV}^{-1}$~s, $t_{\rm ad}\sim R/v=
100\,R_{12}/v_{10}$~s, and $t_{\rm sy}=1/a_{\rm s}B^2E\approx 390\,B_{\rm G}^{-2}\,E_{\rm TeV}^{-1}$~s, 
where $R=10^{12}\,R_{12}$~cm and $v=10^{10}\,v_{10}$~cm~s$^{-1}$ are typical lengths and flow velocities 
in the region, and $D$ the Bohm diffusion coefficient. This yields a maximum energy for Fermi~I acceleration of 
$E_{\rm max}=\min[E_{\rm max}^{\rm diff},E_{\rm max}^{\rm ad},E_{\rm max}^{\rm sy}]$, where $E_{\rm max}^{\rm diff}
\sim 74\,R_{12}\,B_{\rm G}\,(0.5\,c/v_{\rm s})^{-1}$~TeV, $E_{\rm max}^{\rm ad}\sim 36\,B_{\rm G}\,R_{12}\,v_{10}^{-1}\,
\,(0.5\,c/v_{\rm s})^{-2}$ ~TeV, and $E_{\rm max}^{\rm sy}\sim 12\,B_{\rm G}^{-1/2}\,(0.5\,c/v_{\rm s})^{-1}$~TeV. These simple 
estimates already show that for slow shocks (e.g. between two massive star winds), or outside the range 
$B\sim 0.1-1$~G, it is very hard to accelerate particles up to $E\gtrsim 10$~TeV.
It is worth noting that particle acceleration in highly relativistic shocks could work\cite{sir11b}, but requires rather specific
magnetic field conditions downstream, such as high turbulence and relatively small magnetization. This might be hard to
realize in pulsar winds\cite{lem10}, in which a significant toroidal $B$-field is usually expected to remain downstream the
termination shock (see however below).  

In stochastic particle acceleration or Fermi~II, the situation is worse than in Fermi~I. This process is driven only by 
stochastic collisions with randomly moving, magnetic field irregularities. It is a second order process, i.e. $\Delta 
E\propto (v_{\rm A}/c)^2$, where $v_{\rm A}$ is the Alfven speed. Since $t_{\rm acc}\sim (c/v_{\rm A})^2\,E/qBc$,  it is
required that turbulent energy will be a significant fraction of the plasma energy and that $v_{\rm A} \rightarrow  c$. It
means that $B$ should be around equipartition with the turbulence rest mass energy density. Under such a  condition, if
$B\sim 1$~G (i.e. optimal for $v_{\rm A}\sim 0.5\,c$), then $n\sim 100$~cm$^{-3}$, hardly feasible for a compact
binary. Downstream
of a pulsar wind termination shock, relativistic Alfvenic speed may be achieved, although size and turbulence-energy
requirements favor the region behind the pulsar with respect to the star. This however requires negligible stellar wind
contamination and adiabatic losses and may be unrealistic (see Ref. \refcite{bos11}), so a proper assessment of the Fermi~II
efficiency here requires a detailed study. In general, stochastic acceleration  seems to be more suitable to explain extended
and low-energy emission, in regions downstream shocks or rich  in instabilities.

Shear acceleration is, like Fermi~II, a stochastic process, but relies on an additional global velocity gradient in  the flow
$\sim \Delta u/\Delta R$. In the mildly relativistic case, $t_{\rm acc}\sim 3(\Delta R)^2/r_{\rm g}c=300\, B_{\rm G}\,\Delta
R_{11}^2\,E_{\rm TeV}^{-1}$ ($\Delta R= 10^{11}\,\Delta R_{11}$~cm; $r_{\rm g}$ is the particle gyroradius), and therefore
the acceleration  timescale has the same dependence on $E$ as synchrotron. In order to operate then, shear has to overcome
synchrotron cooling, implying $B\lesssim \Delta R_{11}^{-2/3}$~G. Shear also requires of an injection process, since
otherwise adiabatic or advection (escape) losses will block it, i.e. $E>3\,\Delta R_{11}\, v_{10}\,B_G (\Delta
R/R)_{0.1}$~TeV, where $(\Delta R/R)_{0.1}$ means $\Delta R=R/10$. In principle, Fermi~I  could be a good injection
candidate, since it does no require high $B$ to operate. Fermi~II otherwise needs higher $B$ that may render synchrotron
cooling dominant over shear acceleration. The shear maximum  energy (in the mildly relativistic case) is limited by $r_{\rm
g}=\Delta R$, i.e., $E_{\rm max}^{sh}\sim q\,B\,\Delta R \approx 30\,B_{\rm G}\,\Delta R_{11}$~TeV. 

\section{The converter mechanism and magnetic reconnection}

In conventional relativistic ($\Gamma \gg1$) shock scenarios, charged particles are quickly overtaken by the 
shock so that they do not have enough time to isotropise in the upstream region. Thus, when caught up by the 
shock, the shock normal/particle motion angle $\theta$ will be $\sim 2/\Gamma$, and the energy gain, $\Delta 
E/E\sim \Gamma^2\,\theta^2/2$, will be reduced down to $\sim 2$. However, particles may have time to 
isotropise upstream if they managed to 
switch to a neutral state and propagate far from the shock without deflections in the $B$-field. For instance, an electron of energy 
$E$ cooling via Klein-Nishina (KN) inverse Compton (IC) can transfer most of its energy to a scattered photon. 
For pair creation mean free paths $l\sim \Gamma^2\,r_{\rm g}$, the gamma ray will pair create with an ambient 
photon far enough from the shock to allow the subsequent electron (or positron) to get deflected by $\theta\sim 
1$, i.e., $\Delta E/E\sim \Gamma^2/2$. This effectively implies $t_{\rm acc}\sim r_{\rm g}/c$, the electrodynamical 
limit (e.g. Ref. \refcite{aha02}). If otherwise $l$ is too small, 
then $\theta\rightarrow 2/\Gamma$, i.e., the standard case.
Once started, the mechanism proceeds efficiently and yields a rather hard electron spectrum until $l$, 
roughly $\propto E$, becomes $\gtrsim \Gamma\,R$. After that, the electron spectrum becomes 
softer. In
binary systems with bright UV stars, 
the converter mechanism could yield hard electron spectra up to $\sim$~TeV for $\Gamma\sim 100$ 
(see also Ref. \refcite{ste06}). This might be the case in microquasar 
$e^\pm$-jets before suffering mass-loading (caveat: gamma-ray absorption), or at the reaccelerated shocked pulsar wind\cite{bog08}. 
The energy is limited by $r_{\rm g}\sim \Gamma\,R$, although synchrotron losses can reduce this limit. 
For a pulsar 
termination shock in a UV stellar field 
the process cannot work efficiently, since for $\theta\sim 1$, a distance ahead the shock 
of $\Gamma^2\,r_{\rm g}\approx 3\times 10^{21}\,(\Gamma/10^6)^2\,E_{\rm TeV}\,B_{\rm G}^{-1}$~cm would be required. As shear acceleration, 
the converter mechanism requires an injection mechanism. 
In the leptonic case, $E$ should be enough to pair-create in the ambient photon field ($\sim 30$~GeV for stellar
photons). 

Magnetic reconnection is perhaps the less well characterized process among those discussed here. Numerical 
calculations show that, beside bulk acceleration in the current sheet up to $v_{\rm A}$, non-thermal particles can 
be also produced\cite{zen01}. Potentially, the mechanism is fast with $t_{\rm acc}\sim r_{\rm g}/c$ (assuming 
$\epsilon\sim B$, where $\epsilon$ is the current sheet electric field), and particles may reach energies limited 
by $r_{\rm g}\sim\Delta R$, where $\Delta R$ is the current sheet size. However, unless the current sheet
occupies a significant fraction of the source, the process will not yield the required multi-$10$~TeV particle energies. 
A possibility could be many reconnection sites, which could be equivalent to the Fermi~II acceleration process. 
An interesting role of magnetic reconnection may be the dissipation of an alternating polarity $B$-field in the jet 
base (e.g., Ref. \refcite{lyu10}), or at the pulsar wind termination (e.g., Ref. \refcite{lyu08}), thereby 
possibly allowing further acceleration via a Fermi~I-type mechanism in relativistic shocks (e.g., Ref. \refcite{sir11a}). 

\section{Conclusions}

The present work indicates that for typical conditions expected in gamma-ray binaries, the production of photons 
with energies $\gtrsim 10$~TeV indeed requires very efficient acceleration with $t_{\rm acc}< 10-100\,r_{\rm g}/c$. 
Although less strictly, this conclusion also applies to protons. All this implies strong turbulence, and relatively weak
magnetic fields (for electrons), as well as at least mildly relativistic speeds. Fermi~I acceleration, although marginal, 
seems to be the best candidate in mildly relativistic outflows, but for highly relativistic flows the situation is less clear. 
Shear acceleration and Fermi~II could 
also operate, but the latter is unlikely to help at TeV energies. The converter mechanism and magnetic 
reconnection cannot be discarded, although they require quite specific conditions. A better source characterization 
is needed for a proper assessment of the feasibility of all these mechanisms.

\section{Acknowledgments}

We thank Maxim Barkov, Evgeny Derishev and Dmitry Khangulyan for fruitful discussions.
This research has received funding from the European Union Seventh Framework Program (FP7/2007-2013)
under grant agreement PIEF-GA-2009-252463. V.B.-R. acknowledges support by the Spanish Ministerio de Ciencia e Innovaci\'on
(MICINN) under grants AYA2010-21782-C03-01 and FPA2010-22056-C06-02.

\bibliographystyle{ws-procs9x6}
\bibliography{ws-pro-sample}
\end{document}